\documentclass[JIP]{apris}

\usepackage{graphicx}
\usepackage{latexsym}

\usepackage{amsmath}
\usepackage{amsthm}
\usepackage{amssymb}
\usepackage{cite}
\usepackage{cleveref}
\crefname{figure}{Fig.}{Figs.}
\Crefname{figure}{Fig.}{Figs.}

\usepackage[varg]{txfonts}

\usepackage{booktabs}
\usepackage{tabularx}
\usepackage{threeparttable}
\usepackage{float}
\usepackage[linesnumbered,ruled,vlined]{algorithm2e}
\usepackage{algpseudocode}

\usepackage{placeins}

\theoremstyle{definition}
\newtheorem{theorem}{Theorem}
\newtheorem{definition}{Definition}
\newtheorem{corollary}{Corollary}

\usepackage{tikz}
\usepackage{lipsum}

\makeatletter
\def\ieeecopyright{
	\footnotesize © 2025 IPSJ. Personal use of this material is permitted.
	\newline
	DOI: XXX
}
\makeatother
\AddToHook{shipout/firstpage}{
	\begin{tikzpicture}[remember picture,overlay]
		\node[anchor=south west,xshift=1.0cm,yshift=0.8cm] at (current page.south west){
			\parbox{\linewidth}{\raggedright\ieeecopyright}
		};
	\end{tikzpicture}
}

\begin{document}

\title{Accelerating Probabilistic Response-Time Analysis: Revised Critical Instant and Optimized Convolution}

\affiliate{Saitama}{Graduate School of Science and Engineering, Saitama University, Japan}
\affiliate{Academic}{Academic Association (Graduate School of Science and Engineering), Saitama University}
\affiliate{TIERIV}{TIER IV Inc., Japan}

\author{Hiroto Takahashi}{Saitama}[takahashi.h.477@ms.saitama-u.ac.jp]
\author{Atsushi Yano}{Saitama,TIERIV}[]
\author{Takuya Azumi}{Academic,TIERIV}[]

\begin{abstract}
	Accurate estimation of the Worst-Case Deadline Failure Probability (WCDFP) has attracted growing attention as a means to provide safety assurances in complex systems such as robotic platforms and autonomous vehicles. WCDFP quantifies the likelihood of deadline misses under the most pessimistic operating conditions, and safe estimation is essential for dependable real-time applications. However, achieving high accuracy in WCDFP estimation often incurs significant computational cost. Recent studies have revealed that the classical assumption of the critical instant, the activation pattern traditionally considered to trigger the worst-case behavior, can lead to underestimation of WCDFP in probabilistic settings. This observation motivates the use of a revised critical instant formulation that more faithfully captures the true worst-case scenario. This paper investigates convolution-based methods for WCDFP estimation under this revised setting and proposes an optimization technique that accelerates convolution by improving the merge order. Extensive experiments with diverse execution-time distributions demonstrate that the proposed optimized Aggregate Convolution reduces computation time by up to an order of magnitude compared to Sequential Convolution, while retaining accurate and safe-sided WCDFP estimates. These results highlight the potential of the approach to provide both efficiency and reliability in probabilistic timing analysis for safety-critical real-time applications.
\end{abstract}


\maketitle

\section{Introduction}\label{sec: intro}
Dependability in real-time systems is critical for safety-oriented applications such as autonomous vehicles and robotics \cite{Kato2018Autoware,Malik2021Industrial}. In such systems, response-time analysis is fundamental; missing a deadline in hard real-time systems can cause catastrophic failures, whereas soft real-time systems are evaluated through Quality of Service (QoS) metrics \cite{buttazzo2010Soft}.

Traditional analysis relies on the \textit{worst-case execution time} (WCET), which represents the maximum task execution time \cite{Hansen2009Statistical-Based,Puschner2000Guest,Santinelli2014On}. However, WCET-based analysis is often pessimistic for dynamic workloads, such as those in autonomous driving, leading to resource over-provisioning \cite{Abella2015WCET,Wilhelm2008The}. To address this limitation, probabilistic methods modeling execution times as random variables (\textit{probabilistic worst-case execution time}, or pWCET) have gained attention. These methods estimate the \textit{worst-case deadline failure probability} (WCDFP), allowing for the configuration of acceptable deadline miss rates and providing more realistic evaluations, with seminal contributions and rigorous definitions presented in recent literature \cite{roux2021montecarlo,VonDerBruggen2021Efficiently,markovic2022analytical,markovic2021convolution,Bozhko2023What}.

A central challenge in probabilistic response-time analysis is managing the trade-off between estimation accuracy and computational effort. For safety-critical domains, underestimation of the deadline failure probability must be avoided to ensure trustworthy guarantees. Excessive overestimation, however, reduces the practical value of such guarantees. Thus, approximating the true WCDFP with a tight upper bound, computable in a reasonable time, is highly desirable. The classical critical instant, which posits simultaneous task releases to maximize system load, can underestimate the WCDFP in probabilistic analysis, as recent studies show scenarios with higher workloads \cite{liu1989Scheduling,chen2022critical}. Consequently, the redefinition of the critical instant requires a reassessment of analysis methods that relied on the classical assumption.

This paper focuses on circular convolution-based methods for WCDFP estimation, adapting these approaches to the revised critical instant. The main objective is to improve computational efficiency while addressing the accuracy-computation trade-off. For comparison, Monte Carlo-based methods \cite{roux2021montecarlo} and Berry-Esseen inequality-based methods \cite{markovic2022analytical} are also evaluated under the revised critical instant. These evaluations highlight the trade-offs among WCDFP estimation methods in modern probabilistic real-time systems, particularly in robotics and autonomous driving where reliability is paramount.

The main contributions of this paper are:
\begin{itemize}
    \item Introduction of an optimization technique for convolution merge order, enabling efficient WCDFP estimation under the revised critical instant.
    \item A comparison of the optimized convolution method against Sequential Convolution, Monte Carlo, and Berry-Esseen methods, all adapted to the revised critical instant.
    \item Empirical insights into trade-offs between accuracy, execution time, and applicability of WCDFP estimation methods in probabilistic real-time systems.
\end{itemize}

The remainder of this paper is structured as follows. \Cref{sec: model} outlines the system model. \Cref{sec: wcdfp} specifies the definition and upper bound of WCDFP. \Cref{sec: methods} introduces the mathematical and algorithmic foundations of efficient convolution. \Cref{sec: application} presents the proposed accelerated aggregate convolution algorithm incorporating the revised critical instant. \Cref{sec: evaluation} details the evaluation results. \Cref{sec: related work} reviews related work. \Cref{sec: conclusion} concludes the paper and discusses future work.

\section{System Model}\label{sec: model}

In this section, the necessary probabilistic concepts and task set assumptions are introduced, serving as the foundation for the subsequent analysis. The variables and symbols introduced in this section are summarized in \tabref{table:notation}, which provides a comprehensive overview of the notation used throughout the paper.

\subsection{Probabilistic Notation and Symbols}

\begin{table}[tb]
    \centering
    \caption{Summary of notation}
    \label{table:notation}
    \begin{threeparttable}
        \begin{tabularx}{\linewidth}{lX}
            \hline\hline
            \textbf{Symbol}      & \textbf{Explanation}                                                               \\
            \hline
            $\tau$               & A task set.                                                                        \\
            $\tau_i$             & A task from $\tau$ with index $i$.                                                 \\
            $\gamma$             & Minimum indivisible unit of time (e.g., processor cycle).                          \\
            $\mathbb{T}$         & Time domain, $\mathbb{T} = \{ \gamma \cdot k \mid k \in \mathbb{N} \}$.            \\
            $T_i$                & The minimum inter-arrival time of $\tau_i$.                                        \\
            $D_i$                & Relative deadline of $\tau_i$.                                                     \\
            $\pi_i$              & Priority of task $\tau_i$.                                                         \\
            $J_{i,j}$            & The $j$-th job of $\tau_i$.                                                        \\
            $C_i$                & Probabilistic execution time distribution of $\tau_i$.                             \\
            $C_{i,j}$            & Execution time (a specific value) of the $j$-th job $J_{i,j}$ of task $\tau_i$.      \\
            $a_{i,j}$            & Arrival time of $J_{i,j}$.                                                         \\
            $d_{i,j}$            & Absolute deadline of $J_{i,j}$.                                                    \\
            $\Omega$             & Sample space of system evolutions.                                                 \\
            $\omega \in \Omega$  & A sample (particular system evolution) from $\Omega$.                              \\
            $\xi \subseteq \Omega$ & Event encompassing all evolutions exhibiting an identical arrival sequence.      \\
            \hline
        \end{tabularx}
    \end{threeparttable}
\end{table}

The probability space $(\Omega, \mathcal{F}, \mathbb{P})$ is defined as follows: $\Omega$ denotes the set of all possible outcomes (the sample space), $\mathcal{F} \subseteq 2^\Omega$ represents the event space, consisting of subsets of $\Omega$, and $\mathbb{P}: \mathcal{F} \to [0,1]$ is the probability measure. The set of real numbers is denoted by $\mathbb{R}$.

\begin{definition}[Random Variable]
    A random variable $X$ on a probability space $(\Omega, \mathcal{F}, \mathbb{P})$ is defined as a function $X: \Omega \to \mathbb{R}$ such that for any real number $x \in \mathbb{R}$, the set $\{ \omega \in \Omega \mid X(\omega) = x \} \in \mathcal{F}$ is measurable.
\end{definition}

The probability of a random variable $X$ taking a value $x$ is expressed as $\mathbb{P}[\omega \in \Omega \mid X(\omega) = x]$, or in short form, $\mathbb{P}[X = x]$. From this point onward, similar notation is consistently used to represent probabilities and distributions of random variables.

A discrete random variable \(X\) and the corresponding probability mass function (PMF) can be explicitly represented by listing the possible values \(x_k\) and their associated probabilities \(\mathbb{P}[X=x_k]\). Throughout this paper, a two-row matrix notation is adopted for clarity, similar to representations in related literature~\cite{markovic2021convolution}:
\begin{equation}
    \label{eq:rv_representation}
    X \sim \begin{bmatrix} x_1 & x_2 & \dots & x_n \\ \mathbb{P}[X=x_1] & \mathbb{P}[X=x_2] & \dots & \mathbb{P}[X=x_n] \end{bmatrix},
\end{equation}
where \(x_1 < x_2 < \dots < x_n\) are the distinct values \(X\) can take, sorted in increasing order, and \(n\) is the cardinality of the support of \(X\) (the set of values taken by X with positive probability, denoted Im\(X\)).

\begin{definition}[Cumulative Distribution Function (CDF)]
    \label{def:cdf}
    The cumulative distribution function (CDF) of a random variable $X$, denoted by $F_X: \mathbb{R} \rightarrow [0, 1]$, is defined as follows:
    \begin{equation}
        \label{eq:cdf_def}
        F_X(x) \triangleq \mathbb{P}[X \leq x], \quad \forall x \in \mathbb{R}.
    \end{equation}
    The CDF describes the probability that the random variable \(X\) takes on a value less than or equal to \(x\). For a discrete random variable, this function represents the sum of probabilities of all values in the support up to \(x\).
\end{definition}

\begin{definition}[Partial Order of Random Variables]
    \label{def:partial_order_rv}
    For two random variables \( X \) and \( Y \), \( X \) is said to be less than or equal to \( Y \) in distribution, denoted \( X \preceq Y \), if \( F_X(x) \geq F_Y(x) , \forall x \in \mathbb{R} \).
\end{definition}


\subsection{Task Set Assumptions}
\label{ssec:task_set_assumptions}

A system is considered in which time is discretized into multiples of a constant minimum time interval \( \gamma > 0 \), representing the smallest indivisible unit of time, such as a processor cycle. The corresponding time domain is given by \( \mathbb{T} = \{ \gamma \cdot k \mid k \in \mathbb{N} \} \subset \mathbb{R} \). The system follows a fully preemptive fixed-priority scheduling policy, ensuring that the highest-priority ready job is executed at any given time. Within this framework, the system consists of \( n \) sporadic real-time tasks \( \tau \triangleq \{\tau_1, \tau_2, \dots, \tau_n\} \), where each task \( \tau_i \) is assigned a unique priority.

Each task \( \tau_i \) is defined by a tuple \( \left( C_i, T_i, D_i, \pi_i \right) \). The term \(C_i\), referred to in \tabref{table:notation} as the execution time distribution of task \(\tau_i\), represents the random variable for the execution time of task \(\tau_i\) in this paper. This \(C_i\) embodies the probabilistic Worst-Case Execution Time (pWCET) characterization. The concept of pWCET provides an upper probabilistic bound on the inherent, underlying probabilistic execution time (pET) of any job generated by task \(\tau_i\). Let \(\mathcal{C}_i^*\) denote a random variable representing this underlying pET for a job of task \(\tau_i\). The distribution of \(C_i\) is defined such that the distribution stochastically upper-bounds \(\mathcal{C}_i^*\), meaning \( \mathcal{C}_i^* \preceq C_i \) according to \Cref{def:partial_order_rv}. This signifies that \(C_i\) represents a more pessimistic or equally pessimistic execution time profile compared to the underlying pET \(\mathcal{C}_i^*\). The PMF of \(C_i\), as represented by the notation in \Cref{eq:rv_representation}, is assumed to satisfy this pWCET characteristic. A primary motivation for employing such a pWCET abstraction is to facilitate analytical techniques that require random variables to be independent and identically distributed (IID), which is often not true for underlying pETs due to systemic interactions. The pWCET characterization is therefore constructed as a safe, often pessimistic, over-approximation of pETs, specifically to enable this IID assumption for analytical tractability \cite{Bozhko2023What}. Consequently, for many probabilistic analyses, including those in this paper, the pWCET characterizations, \(C_i\), for different tasks \(\tau_i\) and \(\tau_k\) (where \(i \neq k\)) are assumed to be independent random variables. Furthermore, all jobs \(J_{i,j}\) generated from the same task \(\tau_i\) are assumed to have execution times that are IID according to the pWCET distribution \(C_i\).

Following the characterization of task execution times, other task parameters are defined as follows. \( T_i \) is the minimum inter-arrival time, and constrained deadlines are assumed, meaning \( D_i \leq T_i \) for all tasks \( \tau_i \in \tau \). \( D_i \) is the relative deadline, and \( \pi_i \) represents the priority of the task \( \tau_i \), with \( \pi_i > \pi_j \) indicating that \( \tau_i \) has a higher priority than \( \tau_j \). When a task \( \tau_i \) is activated, a job \( J_{i,j} \) is generated, where \( j \) is the job index. Each job \( J_{i,j} \) has a release time \( a_{i,j} \) and must complete execution before the absolute deadline \( d_{i,j} = a_{i,j} + D_i \). If a job misses the deadline, the job is immediately discarded and does not continue executing. The execution time of an individual job \( J_{i,j} \) of task \( \tau_i \), denoted as \(C_{i,j}\), is a specific value sampled from the pWCET distribution \(C_i\). For simplicity, the scenario \( \omega \) is omitted in this notation.

\begin{figure}[!t]
    \centering
    \includegraphics[width=\linewidth]{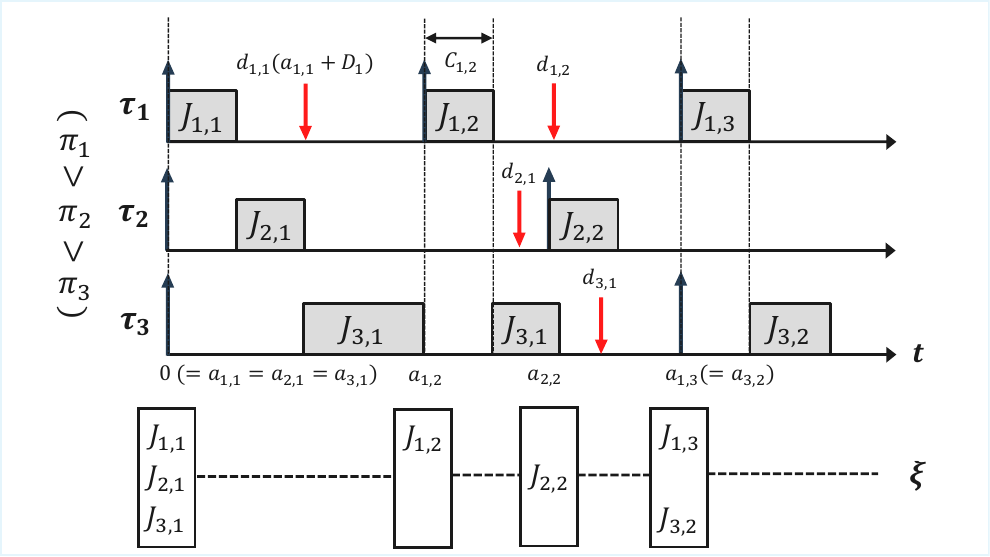} 
    \caption{Visualization of the job arrival sequence, showing jobs released at various time instants for different tasks, along with their respective priorities}
    \label{fig:arrival_sequence}
\end{figure}

The arrival sequence \(\xi\) defines the overall pattern of job arrivals for all tasks over time. Specifically, within a given arrival scenario \(\xi\), the release times, jobs, and absolute deadlines are \(a_{i,j}^\xi\), \(J_{i,j}^\xi\), and \(d_{i,j}^\xi\), respectively, indicating that these values are defined within the context of this particular scenario. As shown in \Cref{fig:arrival_sequence}, this visualization provides an example of job release patterns over time. This foundational framework helps analyze system behavior.

\section{WCDFP and a Revised Upper Bound}\label{sec: wcdfp}

The primary focus of this section is to establish a rigorous foundation for the understanding and estimation of WCDFP. The necessary foundational concepts and notation are defined for the accurate formulation of WCDFP. Following this, the computational challenge of directly calculating WCDFP is addressed, and an upper bound is introduced as a practical alternative.

\subsection{Definitions and Notations}

To define WCDFP rigorously, this subsection introduces foundational concepts and notation. These include workload, carry-in, aborted execution time, and processor demand, essential for accurately modeling response times and deadline failure probabilities. The definitions presented in this subsection follow those established in prior work \cite{markovic2023cta}, ensuring consistency with research.

\begin{definition}[Workload]
    The workload of a task $\tau_i$ over a time interval $[t_1, t_2)$ in arrival sequence $\xi$, denoted by $\mathcal{W}_i^{\xi}[t_1, t_2)$, is defined as the total processing time required by all jobs of $\tau_i$ released in that interval:
    \begin{equation}
        \mathcal{W}_i^{\xi}[t_1, t_2) \triangleq \sum_{j \in \{ j^{\prime} \mid a_{i,j^{\prime}}^{\xi} \in [t_1, t_2) \}} C_{i,j}.
    \end{equation}
\end{definition}

\begin{definition}[Total Workload]
    The total workload of task $\tau_i$ and all higher-priority tasks, including $\tau_i$, over a time interval $[t_1, t_2)$ in arrival sequence $\xi$, denoted by $\mathcal{TW}_i^{\xi}[t_1, t_2)$, is defined as:
    \begin{equation}
        \mathcal{TW}_i^{\xi}[t_1, t_2) \triangleq \sum_{\pi_i \leq \pi_k} \mathcal{W}_k^{\xi}[t_1, t_2).
    \end{equation}
\end{definition}

\begin{definition}[Service Time]
    Let $\sigma(t)$ denote the scheduling function, which returns the job being executed by the processor at each discrete time $t$. The service time of a job $J_{i,j}^{\xi}$ of task $\tau_i$ up to time $t$ in arrival sequence $\xi$, denoted by $\mathcal{S}_{i,j}^{\xi}(t)$, is defined as the total number of discrete time units in the interval $[a_{i,j}^{\xi}, t)$ during which $J_{i,j}^{\xi}$ is included in $\sigma(t')$:
    \begin{equation}
        \mathcal{S}_{i,j}^{\xi}(t) \triangleq \left| \left\{ t' \in [a_{i,j}^{\xi}, t) \cap \mathbb{T} \mid \sigma(t') = J_{i,j}^{\xi} \right\} \right|.
    \end{equation}
\end{definition}

\begin{definition}[Carry-in]
    The carry-in for task $\tau_i$ at time $t$ in arrival sequence $\xi$, denoted by $\mathcal{CI}_{i}^{\xi}(t)$, is defined as:
    \begin{equation}
        \mathcal{CI}_{i}^{\xi}(t) \triangleq
        \begin{cases}
            C_{i,j} - \mathcal{S}_{i,j}^{\xi}(t) & \text{if } \exists j \in \mathbb{N} : a_{i,j}^{\xi} \leq t < d_{i,j}^{\xi}, \\
            0                                    & \text{otherwise}.
        \end{cases}
    \end{equation}
\end{definition}

\begin{definition}[Total Carry-in of Task $\tau_i$]
    The total carry-in for task $\tau_i$ at time $t$ in arrival sequence $\xi$, denoted by $\mathcal{TCI}_{i}^{\xi}(t)$, is defined as the sum of the carry-in contributions from all higher-priority tasks:
    \begin{equation}
        \mathcal{TCI}_i^{\xi}(t) \triangleq \sum_{\pi_i < \pi_k} \mathcal{CI}_{k}^{\xi}(t).
    \end{equation}
\end{definition}

\begin{definition}[Aborted Execution Time]
    The aborted execution time for a task $\tau_i$ at time $t$ in arrival sequence $\xi$, denoted by $\mathcal{KW}_{i}^{\xi}(t)$, represents the remaining workload of a job $J_{i,j}^{\xi}$ of $\tau_i$ that has been preempted and does not complete by the absolute deadline of the job. This metric is defined as:
    \begin{equation}
        \mathcal{KW}_{i}^{\xi}(t) \triangleq
        \begin{cases}
            C_{i,j} - \mathcal{S}_{i,j}^{\xi}(t) & \text{if } \exists j \in \mathbb{N} : d_{i,j}^{\xi} = t, \\
            0                                    & \text{otherwise}.
        \end{cases}
    \end{equation}
\end{definition}

\begin{definition}[Total Aborted Execution Time of Task $\tau_i$]
    The total aborted execution time for task $\tau_i$ and all higher-priority tasks, including $\tau_i$, over a time interval $[t_1, t_2)$ in arrival sequence $\xi$, denoted by $\mathcal{TKW}_i^{\xi}[t_1, t_2)$, is defined as the sum of aborted execution times for all higher-priority tasks, including $\tau_i$:
    \begin{equation}
        \mathcal{TKW}_i^{\xi}[t_1, t_2) \triangleq \sum_{t \in [t_1, t_2)} \sum_{\pi_i \leq \pi_k} \mathcal{KW}_{k}^{\xi}(t).
    \end{equation}
\end{definition}

\begin{definition}[Processor Demand]
    \label{def:PD}
    The processor demand for task $\tau_i$ over an interval $[t_1, t_2)$ in arrival sequence $\xi$, denoted by $\mathcal{E}_i^{\xi}[t_1, t_2)$, is defined as:
    \begin{equation}
        \mathcal{E}_i^{\xi}[t_1, t_2) \triangleq \mathcal{TCI}_i^{\xi}(t_1) + \mathcal{TW}_i^{\xi}[t_1, t_2) - \mathcal{TKW}_i^{\xi}[t_1, t_2).
    \end{equation}
\end{definition}

\begin{definition}[Probabilistic Response Time]
    \label{def:PRT}
    The probabilistic response time \(\mathcal{RT}_{i,j}^{\xi}\) of a job \(J_{i,j}^{\xi}\) of task \(\tau_i\) in arrival sequence \(\xi\) is defined as the earliest time at which the processor demand for \(J_{i,j}^{\xi}\) can be fully met, measured from the release time of the job. Formally, this quantity is given by:
    \begin{equation}
        \mathcal{RT}_{i,j}^{\xi} \triangleq \inf \{ \Delta \mid \Delta > 0 \land \mathcal{E}_i^{\xi}[a_{i,j}^{\xi}, a_{i,j}^{\xi} + \Delta) \leq \Delta \}.
    \end{equation}
\end{definition}

\begin{definition}[Worst Case Deadline Failure Probability (WCDFP)]
    \label{def:WCDFP}
    The Worst Case Deadline Failure Probability (WCDFP) for a task $\tau_i$ is defined as the maximum of the Deadline Failure Probability (DFP) across all possible arrival scenarios $\xi$ and all jobs $J_{i,j}$ of the task, under the worst-case execution scenario $\omega$. Formally, WCDFP is expressed as:
    \begin{equation}
        \text{WCDFP}_i \triangleq \max_{\xi} \max_{j} \mathbb{P}[\mathcal{RT}_{i,j}^{\xi} > D_i].
    \end{equation}
\end{definition}

\subsection{Upper Bound Considering the Critical Instant}
\label{sec:wc_dl_failure_prob_upper_bound}

WCDFP is formally defined as the maximum deadline failure probability across all arrival scenarios. Traditionally, the value was approximated by the critical instant, assuming simultaneous task releases, but probabilistic interpretations reveal this assumption can be insufficient \cite{chen2022critical}. To address this limitation, an indicator \( S_{k,t} \) is introduced to capture both maximum interference from higher-priority tasks and the execution time of \(\tau_k\), enabling feasible WCDFP approximation.

\begin{figure}[t]
    \centering
    \includegraphics[width=\linewidth]{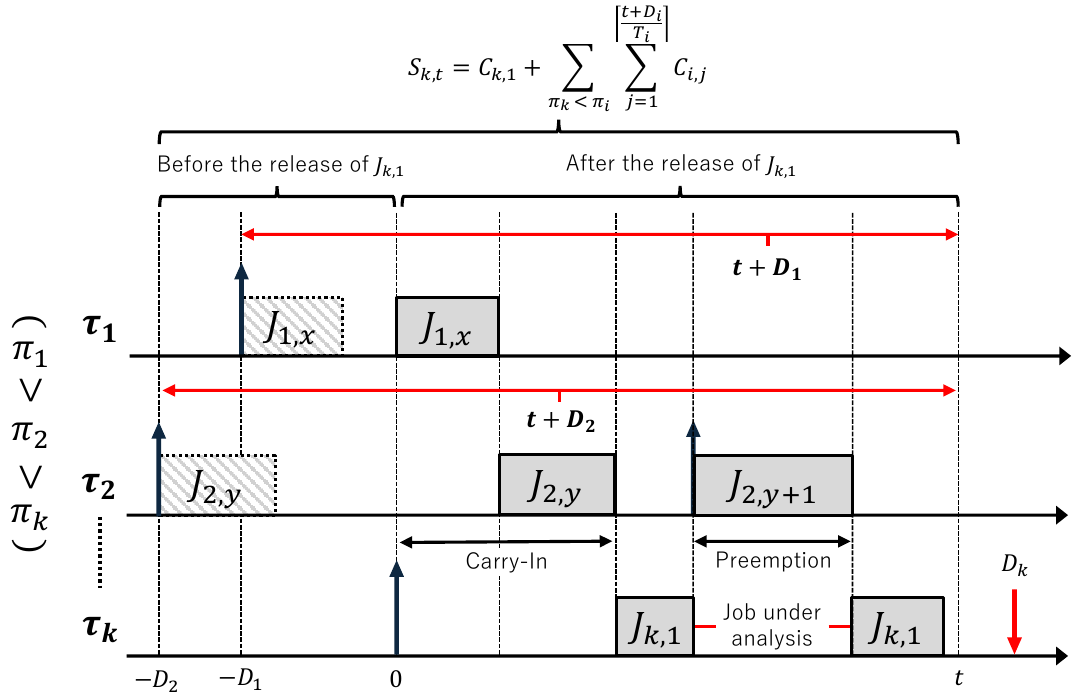}
    \caption{Processor load indicator \( S_{k,t} \) for task \( \tau_k \), including carry-in and preemption effects from higher-priority tasks}
    \label{fig:WCDFP_Skt}
\end{figure}

\begin{corollary}[Corollary 12 in Ref.~\cite{chen2022critical}]
    \label{cor:wcdfp_upper_bound}
    Given a fully preemptive fixed-priority scheduler, a set of constrained-deadline sporadic tasks, and under the assumption that incomplete jobs are aborted at their deadline, the following inequality holds:
    \begin{equation}
        \text{WCDFP}_k \leq \inf_{0 < t \leq D_k} \mathbb{P}(S_{k,t} > t).
    \end{equation}
    Here, \( S_{k,t} \) is defined as:
    \begin{equation}
        S_{k,t} \triangleq C_{k,1} + \sum_{\pi_k < \pi_i} \sum_{j=1}^{\left\lceil \frac{t + D_i}{T_i} \right\rceil} C_{i,j},
    \end{equation}
    where \( \mathbb{P}(S_{k,t} > t) \) represents the probability that the processor demand of task \( \tau_k \) and higher-priority tasks exceeds the time \( t \).
\end{corollary}

The processor load indicator \( S_{k,t} \) is best understood from \Cref{fig:WCDFP_Skt}. The indicator represents the upper bound of processor demand when higher-priority jobs are released with maximum density within \([-D_i, t)\). Under an abort policy where jobs missing their deadlines are terminated, and assuming the analyzed job is released at time 0, the carry-in workload for higher-priority jobs is limited to \([-D_i, 0)\). Unlike the classical assumption of simultaneous releases, the revised indicator allows higher-priority jobs to be released up to \( D_i \) before the analyzed job. This adjustment provides a realistic representation of worst-case interference and enables WCDFP approximation by evaluating \(\mathbb{P}(S_{k,t} > t)\) over \((0, D_k]\), without exhaustively considering all arrival scenarios.

\section{Foundations for Efficient Convolution}\label{sec: methods}

This section overviews convolution-based methods for summing random variables, which form the basis of WCDFP estimation under the revised critical instant. Definitions of independence, convolution, and element-wise product are given, followed by efficient single and repeated convolution techniques.

\subsection{Mathematical Preliminaries for Convolution}
\label{ssec:math_preliminaries_convolution}

This subsection introduces essential mathematical definitions that underpin efficient convolution techniques.

\begin{definition}[Independence of Random Variables]
    \label{def:independence_sec4}
    Two discrete random variables \(X\) and \(Y\) are independent if $\mathbb{P}[X=x, Y=y] = \mathbb{P}[X=x]\mathbb{P}[Y=y]$ for all \(x \in \text{Im}X\) and all \(y \in \text{Im}Y\).
\end{definition}
The independence assumption (\Cref{def:independence_sec4}) is crucial for simplifying the analysis of combined probabilistic behaviors and aligns with pWCET characterization assumptions in \Cref{ssec:task_set_assumptions}.

\begin{definition}[Convolution (Sum of Independent Random Variables)]
    \label{def:convolution_sec4}
    If \(X\) and \(Y\) are independent discrete random variables (\Cref{def:independence_sec4}), their sum \(Z = X+Y\) is also a discrete random variable. The probability mass function (PMF) of \(Z\) is obtained by the linear convolution of the PMFs of \(X\) and \(Y\):
    \begin{equation}
        \label{eq:convolution_def_sec4}
        \mathbb{P}[Z=z] = \sum_{k \in \text{Im}X} \mathbb{P}[X=k]\mathbb{P}[Y=z-k], \quad \forall z \in \text{Im}Z.
    \end{equation}
\end{definition}
The support Im\(Z\) consists of all possible sums of elements from Im\(X\) and Im\(Y\). Direct computation has \(\mathcal{O}(NM)\) complexity, with support size up to \(N+M-1\). Repeated convolution causes quadratic growth and explosion, hindering large-scale analysis.

\begin{definition}[Element-wise Product]
    \label{def:element_wise_product_sec4}
    The element-wise product of two vectors \(A, B\) of length \(n\), denoted \(A \odot B\), results in vector \(C = A \odot B\) of length \(n\), where each element \(c_i = a_i \cdot b_i\).
\end{definition}
The element-wise product (\Cref{def:element_wise_product_sec4}) is central to efficient frequency-domain convolution (\Cref{ssec:efficient_convolution_techniques}).

\subsection{Efficient Convolution Techniques}
\label{ssec:efficient_convolution_techniques}

The \(\mathcal{O}(NM)\) complexity of direct linear convolution (\Cref{def:convolution_sec4}) and associated state-space expansion are major bottlenecks. This subsection details two techniques to mitigate this: Fast Fourier Transform (FFT) for single convolutions and exponentiation by squaring for repeated convolutions.

\subsubsection{Single Convolution using Fast Fourier Transform}
\label{sssec:fft_convolution_detail}

The Fast Fourier Transform (FFT) \cite{Cooley1965An} substantially reduces convolution complexity, based on the Convolution Theorem \cite{Papoulis1962}.

\begin{theorem}[Convolution Theorem \cite{Papoulis1962}]
    \label{thm:convolution_theorem}
    For discrete sequences \(V_X, V_Y\), the Discrete Fourier Transform (DFT) of their linear convolution (\(V_X * V_Y\)) equals the element-wise product (\Cref{def:element_wise_product_sec4}) of their DFTs, assuming appropriate zero-padding of \(V_X, V_Y\):
    \begin{equation}
        \mathcal{F}\{V_X * V_Y\} = \mathcal{F}\{V_X'\} \odot \mathcal{F}\{V_Y'\},
    \end{equation}
    where \(\mathcal{F}\) is the DFT operation.
\end{theorem}
As described in \Cref{thm:convolution_theorem}, time-domain convolution becomes a frequency-domain element-wise product. To compute \(Z=X+Y\), PMFs \(V_X, V_Y\) are zero-padded to \(L\), transformed via FFT, multiplied element-wise, and converted back with an inverse FFT. With FFT/IFFT complexity of \(\mathcal{O}(L \log L)\), overall complexity is \(\mathcal{O}(L \log L)\). This enables larger supports within practical time limits \cite{markovic2021convolution}. However, FFT-based convolution requires memory proportional to \(L\) and can suffer from numerical errors due to floating-point arithmetic, especially for distributions with very small probabilities or a wide range of probability values.

\subsubsection{Exponentiation by Squaring in Convolutions}
\label{sssec:exponentiation_by_squaring_detail}

For \(k\) IID variables, exponentiation by squaring exploits associativity to reduce the number of convolutions to \(\mathcal{O}(\log k)\) \cite{Gordon1998A}. The idea is to compute PMFs for sums corresponding to powers of two (\(S_1, S_2=S_1*S_1, S_4=S_2*S_2,\dots\)) using FFT-based convolution (\Cref{sssec:fft_convolution_detail}). For \(k\) instances, the binary representation of \(k\) determines which \(S_{2^j}\) PMFs are convolved. For example, \(13=8+4+1\) requires \(S_8, S_4, S_1\). Each of these \(\mathcal{O}(\log k)\) convolutions uses the FFT-based method. This efficiency is beneficial for analyzing execution times of multiple jobs from a single task, assuming IID execution times.

\section{Accelerated Aggregate Convolution with Revised Critical Instant}\label{sec: application}

This section details the proposed \textbf{Accelerated Aggregate Convolution} for efficient WCDFP estimation, integrating the revised critical instant detailed in \Cref{sec:wc_dl_failure_prob_upper_bound}. The theoretical basis for optimizing the convolution merge order is explored, contrasting naive sequential merging with a Huffman-based strategy, before presenting the pseudocode of the algorithm.

\subsection{Optimizing Convolution Merge Order}
\label{ssec:optimizing_merge_order}

When computing the probabilistic distribution resulting from convolving multiple independent random variables, such as the execution times of numerous jobs, the order in which these convolutions are performed significantly impacts the overall computational cost. The Aggregate Convolution approach, central to this work, combines these distributions. While the efficiency of individual convolution operations is addressed by techniques like the FFT, detailed in \Cref{sec: methods}, this subsection focuses on a complementary optimization: minimizing the sum of intermediate distribution sizes encountered during the merging process.

\begin{table}[t]
    \centering
    \caption{Helper Functions for Accelerated Aggregate Convolution}
    \label{table:helper_functions_acc_agg_conv}
    \begin{threeparttable}
        \begin{tabularx}{\linewidth}{lX}
            \hline\hline
            \textbf{Function Signature}                & \textbf{Description}                                                                                                                                                \\
            \hline
            \textnormal{CircularConvolution}(\(a, b\)) & Computes the discrete convolution of two PMF vectors \(a\) and \(b\) using FFT for efficient computation.                                                           \\
            \textnormal{TruncateAndSum}(\(a, [b, c)\)) & Truncates PMF vector \(a\) to range \([b, c)\) by removing elements outside this interval. Returns the sum of probabilities of the removed (out-of-range) elements. \\
            \hline
        \end{tabularx}
    \end{threeparttable}
\end{table}

\subsubsection{Worst-Case Cost of Sequential Merging}
\label{sssec:worst_case_sequential_merging_app}

Consider \(N\) initial PMFs with sizes \(w_1, \ldots, w_N\) (total size \(S = \sum w_i\)). Sequential convolution (e.g., \((V_1*V_2)*V_3 \ldots\)) causes progressive growth of intermediate distribution sizes. The cost, often measured by the sum of all input distribution sizes during merging, can be significant. For a sequential merge, the sum of operand sizes over \(N-1\) convolutions can reach \(\mathcal{O}(SN)\) in the worst case. This occurs if large distributions form early and are repeatedly convolved, posing challenges for large \(N\) or \(S\).

\subsubsection{Huffman-Based Optimal Merging Strategy}
\label{sssec:huffman_merging_app}

A more efficient strategy for merging \(N\) distributions aims to minimize the total computational effort by carefully selecting the pair of distributions to convolve at each step. This problem is analogous to constructing an optimal prefix code, a task solved by Huffman's algorithm \cite{huffman1952method}. In this analogy, the initial PMF vectors correspond to symbols, and the size of each vector (\(w_i\)) corresponds to the frequency of a symbol. The merging process is equivalent to building a Huffman tree, where each internal node represents the convolution of the two child nodes.

Huffman's algorithm constructs this tree by iteratively merging the two nodes with the smallest current weights (sizes). This process is known to minimize the total weighted external path length, \(\sum_{i=1}^{N} w_i d_i\), where \(d_i\) is the depth of a leaf. In the context of convolution, minimizing this sum effectively minimizes the total sum of operand sizes across all steps. By always merging the two smallest available distributions, the Huffman-based strategy ensures that larger distributions are, on average, involved in fewer merging steps. While \(N-1\) convolutions are always needed, the Huffman strategy minimizes the sum of operand sizes for these operations to a bound of \(\mathcal{O}(S \log N)\) \cite{huffman1952method}, significantly improving upon the naive sequential \(\mathcal{O}(SN)\) worst-case bound. This merge sequence optimization complements other techniques such as FFT-based convolution and exponentiation by squaring.

\subsection{Implementation of Accelerated Aggregate Convolution}
\label{ssec:implementation_accelerated_aggregate_convolution_final}

\begin{algorithm}[t]
    \SetAlgoVlined
    \caption{Accelerated Aggregate Convolution for WCDFP}
    \label{alg:accelerated_aggregate_convolution}
    \KwIn{\(\tau\), \(P_C\) (map of pWCET PMFs \(P_{C_i}\)), \(J_{x,y}\) (Target Job)}
    \KwOut{WCDFP for \(J_{x,y}\)}

    \(\textnormal{pdf\_array} \gets \emptyset\)\;
    \(\textnormal{WCDFP} \gets 0.0\)\;
    \(\textnormal{priority\_queue} \gets \emptyset\) \tcp{Min-queue for (size, index)}

    \For{\( \tau_i \in \{ \tau_{j} \mid \pi_{j} \geq \pi_{x} \}\)}{
        \(\textnormal{release\_count} \gets \left\lceil \dfrac{d_{x,y} + D_i}{T_i} \right\rceil\)\;
        \If{\(i = x\)}{ 
            \(\textnormal{release\_count} \gets 1\)\;
        }

        \(\textnormal{pdf} \gets P_{C_i}\)\;
        \While{\(\textnormal{release\_count} > 0\)}{
            \If{\(\textnormal{release\_count} \bmod 2 = 1\)}{
                \(\textnormal{idx} \gets \textnormal{length}(\textnormal{pdf\_array})\)\;
                \(\textnormal{pdf\_array} \gets \textnormal{pdf\_array} \cup \{\textnormal{pdf}\}\)\;
                \(\textnormal{priority\_queue.push}\bigl(\,(\textnormal{length}(\textnormal{pdf}),\, \textnormal{idx})\bigr)\)\;
            }
            \(\textnormal{pdf} \gets \textnormal{CircularConvolution}(\textnormal{pdf}, \textnormal{pdf})\)\;
            \(\textnormal{release\_count} \gets \left\lfloor \dfrac{\textnormal{release\_count}}{2} \right\rfloor\)\;
        }
    }

    \While{$\mathrm{length(priority\_queue)} > 1$}{
        $(\textnormal{size}1, \textnormal{idx}1) \gets \textnormal{priority\_queue.pop}()$\;
        $(\textnormal{size}2, \textnormal{idx}2) \gets \textnormal{priority\_queue.pop}()$\;
        $\textnormal{merged\_pdf} \gets \textnormal{CircularConvolution}
            \bigl(\textnormal{pdf\_array}[\,\textnormal{idx}1\,], \textnormal{pdf\_array}[\,\textnormal{idx}2\,]\bigr)$\;

        $\mathrm{WCDFP} \gets \mathrm{WCDFP} + \textnormal{TruncateAndSum}\bigl(\textnormal{merged\_pdf}, [\,0, d_{x,y}\,)\bigr)$\;

        $\textnormal{pdf\_array}[\,\textnormal{idx}1\,] \gets \textnormal{merged\_pdf}$\;
        $\textnormal{priority\_queue.push}\bigl(
            \textnormal{length}(\textnormal{merged\_pdf}),\, \textnormal{idx}1
            \bigr)$\;
    }

    \Return \(\textnormal{WCDFP}\)\;
\end{algorithm}

The \textbf{Accelerated Aggregate Convolution} method for WCDFP estimation incorporates the revised critical instant to identify relevant job releases and employs the Huffman-based merging strategy for efficient convolution of the resulting PMFs. The key helper functions in this algorithm are summarized in \tabref{table:helper_functions_acc_agg_conv}.

As shown in Algorithm~\ref{alg:accelerated_aggregate_convolution}, the core logic first involves generating per-task workload PMFs. For each interfering task \(\tau_i\), the algorithm determines the number of relevant job releases (\(\textnormal{release\_count}\)) under the revised critical instant (Lines 4-6). This count considers potential carry-in effects and concurrent execution; for the task \(\tau_x\) under analysis, the count is one. Then, using the pWCET distribution \(P_{C_i}\), the aggregate PMF for these jobs is computed efficiently via exponentiation by squaring (Lines 9-15). The resulting PMFs are collected in \(\textnormal{pdf\_array}\) with references (size and index) pushed to a \(\textnormal{priority\_queue}\).

With all PMF components in the \(\textnormal{priority\_queue}\), a Huffman-like merge occurs (Lines 16-25). The two smallest PMFs are repeatedly extracted and convolved via \textnormal{CircularConvolution}. The sum of probabilities outside the deadline, obtained from the \textnormal{TruncateAndSum} function, is added to the \textnormal{WCDFP}. The resulting PMF is reinserted into the \(\textnormal{priority\_queue}\). This procedure continues until one PMF remains, returning the accumulated \textnormal{WCDFP}.

\section{Evaluation}\label{sec: evaluation}

This section presents the evaluation of the methods, focusing on their effectiveness and computational efficiency under various task set configurations.

All evaluated methods incorporate the revised critical instant, a key aspect of this work, addressing recent findings that classical assumptions can lead to WCDFP underestimation \cite{chen2022critical}. Throughout this evaluation, four approaches are compared: the Monte Carlo method (\textbf{MC}); Sequential Convolution (\textbf{SC}), which performs convolutions iteratively following the arrival sequence $\xi$, processing the pWCET of one job at a time; Aggregate Convolution (\textbf{AC}), with two variants: \textbf{AC (Orig.)}, an unoptimized version that applies exponentiation by squaring to convolve multiple jobs from the same task and then merges distributions from different tasks without considering specific arrival order, and \textbf{AC (Imp.)}, the optimized version that incorporates Huffman-based convolution merge order optimization \cite{huffman1952method} for enhanced speed when combining multiple distributions; and the Berry-Esseen theorem-based approximation (\textbf{BE}). SC, AC (Orig.), and AC (Imp.) are convolution-based methods, with SC serving as a baseline for high accuracy and AC (Imp.) being the primary contribution regarding performance optimization.

\subsection{Experiment Setup}

This evaluation considers 1,500 sporadic task sets, evenly distributed across 10 task cardinalities \{10, 20, 30, 40, 50, 60, 70, 80, 90, 100\} and three utilization levels \{0.60, 0.65, 0.70\}. For each combination of cardinality and utilization, 50 task sets were generated. The inter-arrival time \( T_i \) of task \( \tau_i \) was sampled logarithmically within the range [10 ms, 1000 ms]. Each task \( \tau_i \) has a utilization \( U_i \) and a WCET denoted by \( \mathrm{WCET}_i \). The utilization \( U_i \) was assigned using the Dirichlet-Rescale algorithm to balance workload distribution across tasks \cite{Griffin2020Generating}. Execution times follow a truncated mixture distribution. Under normal operation, the execution time distribution is modeled as \( \mathcal{N}(\mu_i, \sigma_i) \), where \( \mu_i = \mathrm{WCET}_i / 3.0 \) and \( \sigma_i = \mathrm{WCET}_i / 6.0 \). For abnormal operation, the execution time is modeled as \( \mathcal{N}(\mu_i', \sigma_i') \), where \( \mu_i' = \mathrm{WCET}_i / 1.2 \) and \( \sigma_i' = \mathrm{WCET}_i / 30.0 \). The final execution time is defined as a weighted mixture of these two distributions:
\[
    \tilde{C}_i \sim 0.95 \cdot \mathcal{N}(\mu_i, \sigma_i) + 0.05 \cdot \mathcal{N}(\mu_i', \sigma_i')
\]
truncated to the range [0, \( \mathrm{WCET}_i \)], then discretized to the nearest unit \( \gamma \) and normalized, yielding the final distribution \( C_i \). These distributions were generated based on empirical observations of Autoware runtimes measured by CARET \cite{Kuboichi2022CARET,Toba2024Deadline}. The WCET values were uniquely derived based on the assigned utilization \( U_i \), the task periods \( T_i \), and the execution time distribution \( C_i \), ensuring consistency across all task sets. Tasks were prioritized using rate-monotonic scheduling, and the minimum time unit \( \gamma \) was set to 1 \textmu s to balance computational efficiency with precision. The task model assumes constrained deadlines, defined as \( D_i \leq T_i \) in \figref{sec: model}. For simplicity, this evaluation assumes \( D_i = T_i \).

Monte Carlo (\textbf{MC}) simulations were performed with a fixed sample size of 100,000. For \textbf{SC} and Aggregate Convolution (\textbf{AC}), Python implementations used the \texttt{fftconvolve} function from the \texttt{scipy} library to compute circular convolutions efficiently. Berry-Esseen (\textbf{BE}) was evaluated using the same parameters as specified in previous studies, ensuring consistency with existing approaches (e.g., \cite{markovic2022analytical}). The simulations were executed on an AMD Ryzen Threadripper 7960X processor (24 cores, 48 threads) with 128 MiB of L3 cache, running Ubuntu 22.04 LTS. Most methods were implemented in Python and executed in a single-threaded mode to ensure fair performance comparisons. In contrast, \textbf{MC} utilized 24 threads for parallel sample generation, taking advantage of the independence of these samples to reduce runtime without sacrificing accuracy. Results were aggregated at the end of computation, and this parallel execution must be taken into account when interpreting runtime comparisons.

\subsection{Efficiency and Accuracy in WCDFP Estimation}

\begin{figure}[!t]
    \centering
    \includegraphics[width=\linewidth]{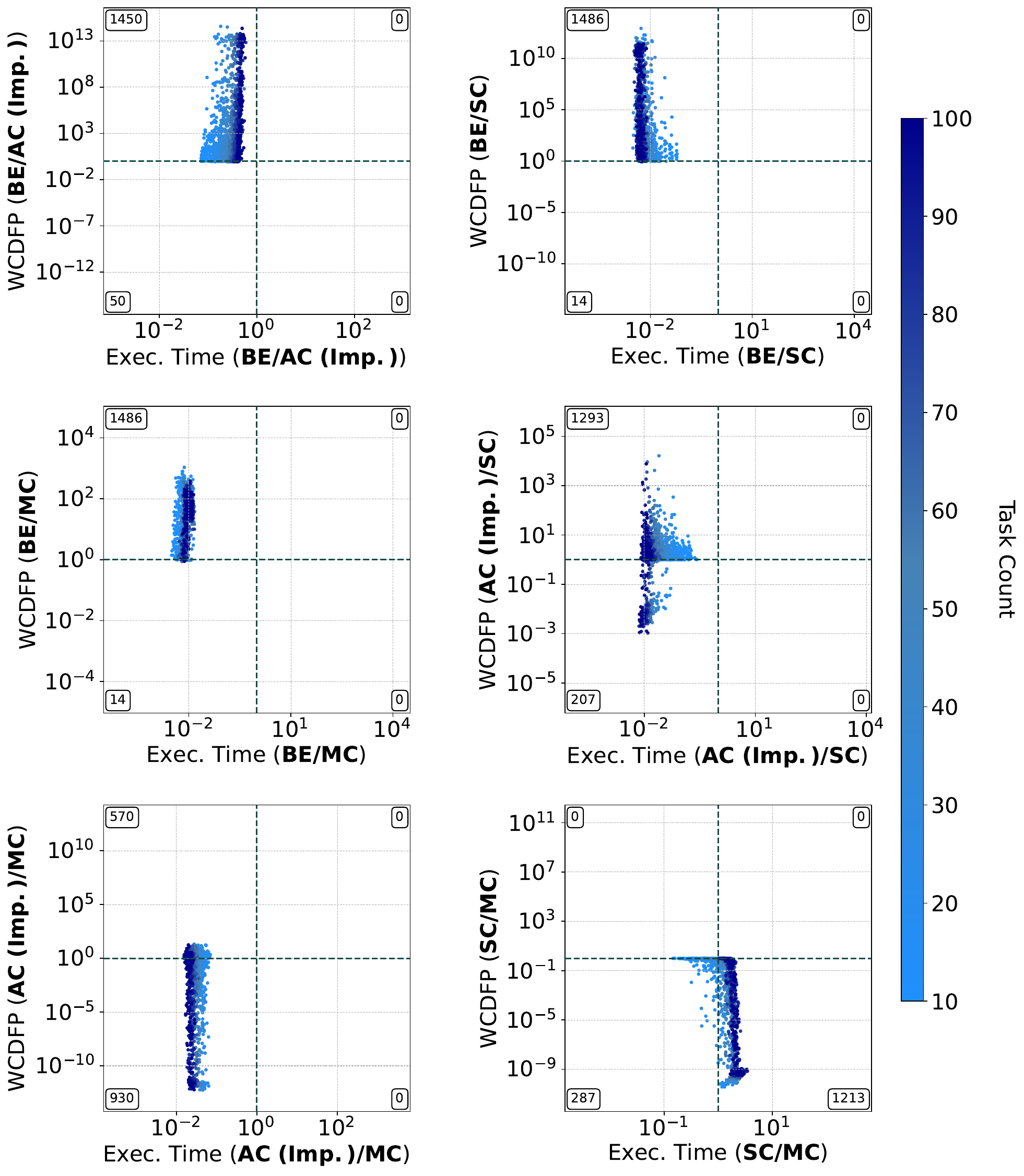}
    \caption{Comparison of WCDFP estimation accuracy and execution time across \textbf{MC}, \textbf{SC}, \textbf{AC (Imp.)}, and \textbf{BE}. Points are positioned relative to baseline values at $x=1.0$ and $y=1.0$, with quadrant counts displayed in the corners}
    \label{fig:ratio_comparison}
\end{figure}

This subsection examines the relationship between execution time and WCDFP estimation accuracy across the four evaluated methods, with a particular focus on the performance of the convolution-based approaches. Comparisons illustrated in \figref{fig:ratio_comparison} show the relative differences between methods. Each point represents the ratios of accuracy and execution time, measured relative to baseline values at $x=1.0$ and $y=1.0$.

The \textbf{BE} method, while consistently achieving the fastest execution times (10 to 100 times faster than other methods), generally yielded less accurate WCDFP estimates, especially against \textbf{SC} and \textbf{AC (Imp.)}.
The \textbf{MC} method was more accurate than \textbf{BE}. However, when compared against the convolution methods, \textbf{MC} showed lower accuracy in many cases (e.g., 930 cases against \textbf{AC (Imp.)}) and did not outperform \textbf{SC} in any evaluated scenario for the given sample size. In terms of execution time, \textbf{MC} required approximately 10 times more computation than \textbf{AC (Imp.)}.

\begin{figure}[!t]
    \centering
    \includegraphics[width=\linewidth]{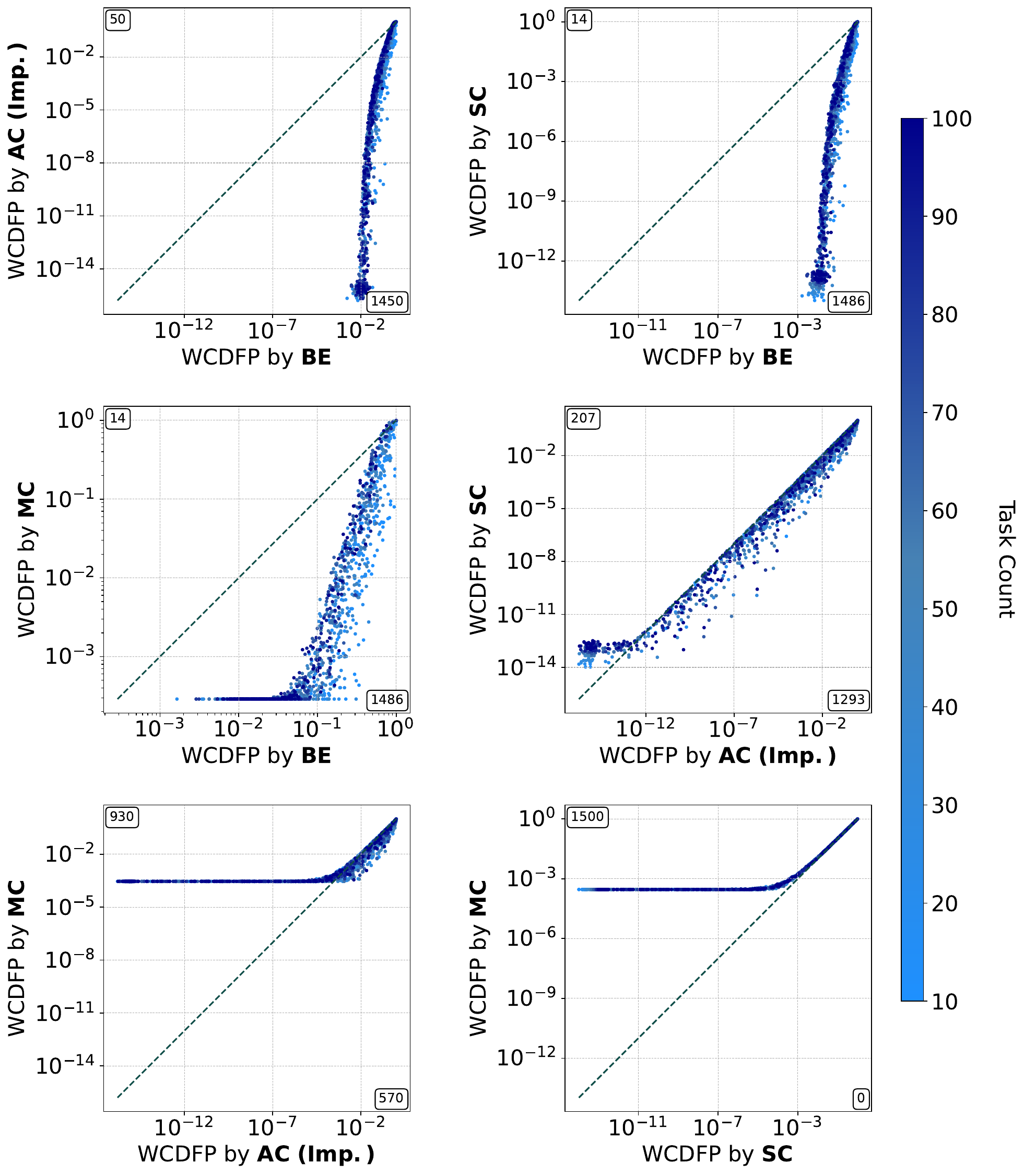}
    \caption{Comparison of WCDFP values computed using \textbf{MC}, \textbf{SC}, \textbf{AC (Imp.)}, and \textbf{BE}. Each point represents a taskset, with counts of task sets above and below the $y = x$ line displayed in the top-left and bottom-right corners, respectively}
    \label{fig:wcdfp_comparison}
\end{figure}

The primary comparison between the convolution methods, \textbf{SC} and \textbf{AC (Imp.)}, revealed that while \textbf{AC (Imp.)} was consistently faster, \textbf{SC} was generally more accurate. This behavior is detailed in \figref{fig:ratio_comparison}. Interestingly, in 207 cases, \textbf{AC (Imp.)} showed better accuracy than \textbf{SC}. This phenomenon, often observed when the true WCDFP is extremely low, is attributed to numerical errors in circular convolution; such errors are less likely to propagate extensively in \textbf{AC (Imp.)} due to the reduced number of convolution operations of \textbf{AC (Imp.)}. Additional details regarding absolute accuracy comparisons are presented in \figref{fig:wcdfp_comparison}.

\begin{figure}[!t]
    \centering
    \includegraphics[width=0.8\linewidth]{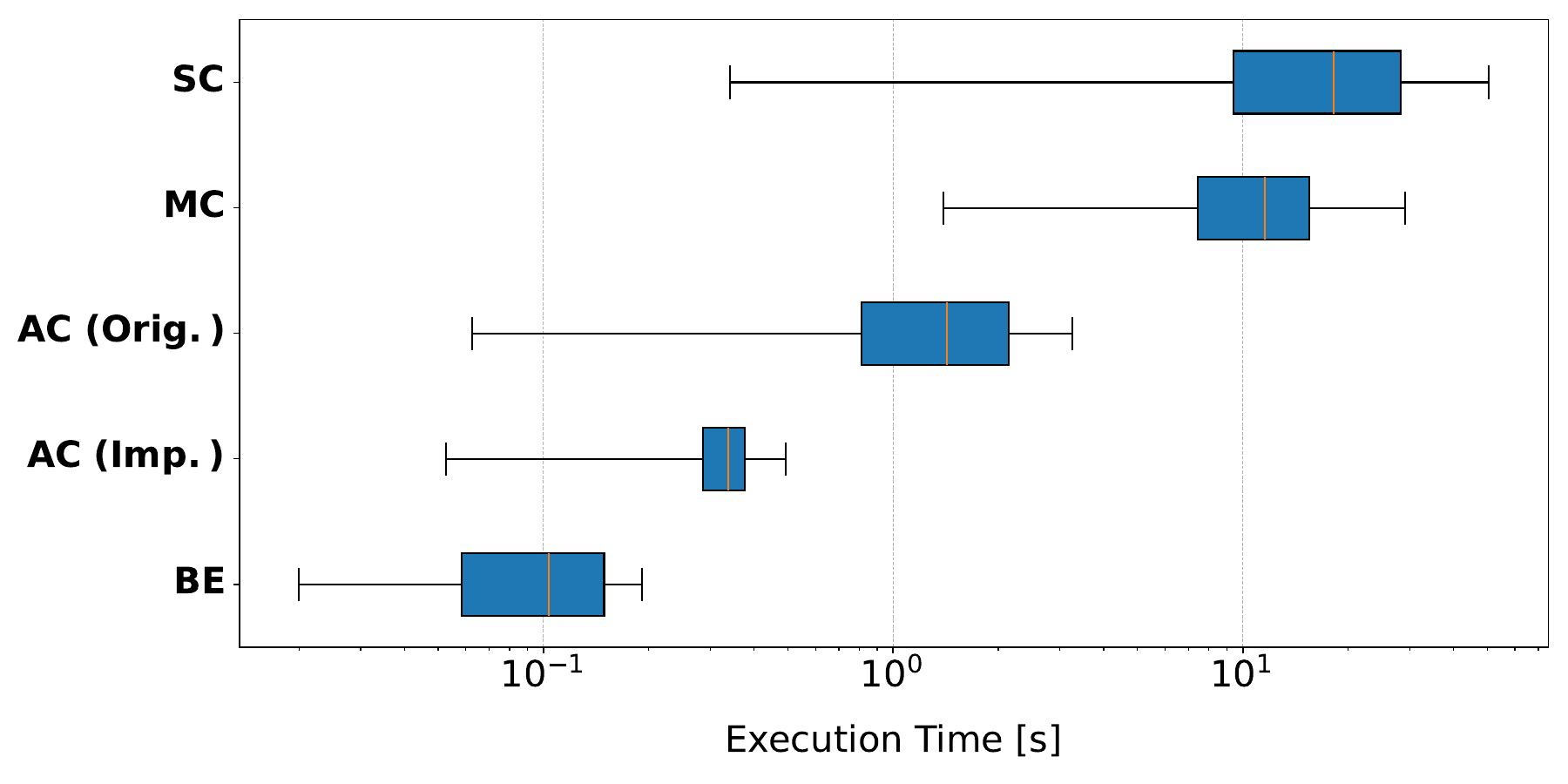}
    \caption{Comparison of execution times for WCDFP estimation across \textbf{MC}, \textbf{SC}, \textbf{AC (Orig.)}, \textbf{AC (Imp.)}, and \textbf{BE}. The boxplots summarize execution time distributions, with each method corresponding to a box}
    \label{fig:time_comparison}
\end{figure}

\subsection{Detailed Accuracy and Performance Analysis}
\label{subsec:detailed_analysis_convolution}

This subsection further examines the accuracy of WCDFP estimates and the execution times of \textbf{SC} and \textbf{AC (Imp.)}, emphasizing the impact of the proposed optimization in \textbf{AC (Imp.)}. Data for MC and BE from Figs.~\ref{fig:wcdfp_comparison} and \ref{fig:time_comparison} serve as context.

For WCDFP values exceeding $10^{-12}$, \textbf{SC} achieved up to four orders of magnitude higher precision than \textbf{AC (Imp.)}, underscoring the suitability of \textbf{SC} for scenarios demanding the highest precision, albeit at the cost of significantly longer execution times (tens of times slower than \textbf{AC (Imp.)}, as shown in \figref{fig:time_comparison}). Conversely, for WCDFP values below $10^{-12}$, \textbf{AC (Imp.)} occasionally outperformed \textbf{SC} in accuracy. This behavior is attributed to \textbf{AC (Imp.)} performing fewer convolution operations, which reduces the accumulation of numerical errors, a critical factor when dealing with extremely small probabilities.

The core contribution related to computational efficiency, the optimization implemented in \textbf{AC (Imp.)}, yielded substantial performance improvements. As shown in \figref{fig:time_comparison}, \textbf{AC (Imp.)} demonstrated an average speedup of about 5 times over \textbf{AC (Orig.)}. This speedup enabled \textbf{AC (Imp.)} to complete WCDFP analyses within 1 second for many configurations, making advanced convolution-based WCDFP estimation under the revised critical instant significantly more practical. While this optimization primarily targets the order of convolution operations, and thus theoretical precision differences between \textbf{AC (Imp.)} and \textbf{AC (Orig.)} are not the central concern, the reduction in the size of intermediate distributions handled during the merging process in \textbf{AC (Imp.)} can potentially contribute to a reduction in absolute numerical error.

For context, \textbf{MC} estimates, with a sample size of 100,000, converged near $10^{-4}$ and did not match the precision of \textbf{SC} (see \figref{fig:wcdfp_comparison}).
The \textbf{BE} method, while extremely fast (average $10^{-1}$ seconds, as shown in \figref{fig:time_comparison}), exhibited clear precision limitations, especially for WCDFP values below $10^{-3}$ (see \figref{fig:wcdfp_comparison}).

\section{Related Work}\label{sec: related work}
This section presents an overview of existing studies and their differences from this paper, with a summary shown in \tabref{table:comparison}.

Recent research has redefined the critical instant for probabilistic response-time analysis, showing that simultaneous task releases do not always yield the worst-case scenario~\cite{chen2022critical}. This finding underscores the need for methods integrating such insights, as classical assumptions may lead to WCDFP underestimation.

Convolution-based methods~\cite{markovic2021convolution} approximate response-time distributions by convolving task execution time distributions. While capable of precise results, scalability can be limited by computational costs, and many existing approaches rely on the classical critical instant assumption. To improve efficiency, advancements like circular convolution and down-sampling have been introduced. In contrast to prior work using coarser discretizations (e.g., 50 µs \cite{markovic2021convolution}), this paper employs a finer 1µs discretization for higher resolution. This work builds upon these foundational techniques by applying the revised critical instant and introducing further optimizations for aggregate convolution.

Monte Carlo-based analysis~\cite{roux2021montecarlo} estimates WCDFP via random sampling, offering scalability, particularly for systems with many tasks. However, this method often relies on the classical critical instant and can require substantial samples to achieve high precision for very low WCDFP values. Notably, while some evaluations use simplified execution time models (e.g., bimodal distributions \cite{roux2021montecarlo}), this paper applies MC analysis to the same high-resolution and realistic mixture model distributions used for all evaluated methods, ensuring a consistent comparison.

The Berry-Esseen inequality-based analysis~\cite{markovic2022analytical} offers an analytical approximation using statistical properties via the Lyapunov central limit theorem. This approach is computationally efficient and avoids the combinatorial explosion of convolutions, though the bound tightness can vary as the method abstracts away fine-grained distribution details.

Cantelli inequality-based methods address WCDFP estimation for dependent tasks. Correlation Tolerant Analysis (CTA)~\cite{markovic2023cta} provides bounds from mean and standard deviation but can be pessimistic due to not explicitly considering task dependencies. Correlation Aware Analysis (CAA)~\cite{markovic2024caa} improves upon CTA by incorporating covariance bounds for more accurate estimates.

\begin{table}[!t]
    \centering
    \caption{Comparison of related methods and features}
    \label{table:comparison}
    \begin{threeparttable}
        \begin{tabularx}{\linewidth}{l>{\centering\arraybackslash}X>{\centering\arraybackslash}X>{\centering\arraybackslash}X>{\centering\arraybackslash}X>{\centering\arraybackslash}X>{\centering\arraybackslash}X}
            \hline\hline
            \textbf{}                                 & \textbf{RCI} & \textbf{BE} & \textbf{MC} & \textbf{CC} & \textbf{ACC} & \textbf{HRED} \\
            \hline
            RTSS 2023~\cite{markovic2023cta}          & \checkmark   & \checkmark  &             &             &              &               \\
            RTSS 2024~\cite{markovic2024caa}          & \checkmark   & \checkmark  &             &             &              &               \\
            RTSS 2021~\cite{roux2021montecarlo}       &              &             & \checkmark  &             &              &               \\
            ECRTS 2021~\cite{markovic2021convolution} &              &             &             & \checkmark  &              &               \\
            RTSS 2022~\cite{markovic2022analytical}   & \checkmark   & \checkmark  &             & \checkmark  &              &               \\
            This paper                                & \checkmark   & \checkmark  & \checkmark  & \checkmark  & \checkmark   & \checkmark    \\
            \hline
        \end{tabularx}
    \end{threeparttable}

    \begin{flushleft}
        \begin{tabular}{ll}
            RCI:  & Consideration of Revised Critical Instant                \\
            BE:   & Berry-Esseen Correction (included as a baseline method)  \\
            MC:   & Monte Carlo Simulation (included as a baseline method)   \\
            CC:   & Circular Convolution (included as a baseline method)     \\
            ACC:  & Accelerated Circular Convolution with Order Optimization \\
            HRED: & High-Resolution and Realistic Execution Distributions    \\
        \end{tabular}
    \end{flushleft}
\end{table}

\section{Conclusions}\label{sec: conclusion}

This paper enhanced WCDFP estimation by incorporating the revised critical instant into convolution-based analysis and introducing an optimized Aggregate Convolution technique. The Huffman-based merge ordering strategy significantly improved efficiency while preserving the safe-sided accuracy required for dependable real-time systems. Evaluations with realistic execution-time distributions confirmed that the optimized method outperforms Sequential Convolution in computation time, while still providing reliable WCDFP estimates. The study emphasized that method selection requires balancing accuracy, cost, and target WCDFP characteristics, with the revised critical instant as a key assumption for trustworthy analysis.

Future work includes extending the method to systems with task dependencies, exploring sustained high-utilization scenarios that challenge current interference assumptions, and investigating parallel implementations to improve scalability. Ultimately, advancing robustness and broader adoption of accurate WCDFP estimation techniques remains a vital pursuit in real-time systems.

\begin{acknowledgment}
	This work was supported by JST AIP Acceleration Research JPMJCR25U1 and JST CREST Grant Number JPMJCR23M1, Japan.
\end{acknowledgment}

\bibliographystyle{ipsjunsrt-e}
\bibliography{ref.bib}

\end{document}